\documentclass[superscriptaddress,prl,twocolumn,showpacs,preprintnumbers,altaffilletter,aps,showpacs,nofootinbib]{revtex4-1}
\usepackage{color}      
\usepackage{graphicx}
\usepackage{subfigure}
\usepackage{multirow}
\usepackage{siunitx} 
\usepackage{datetime} 
\usepackage{xspace}
\usepackage{hyphenat}
\usepackage{lineno}

\newcommand{\isotope}[2]{$^{#2}{\rm #1}$\xspace}
\newcommand{\me}{\ensuremath{m_\mathrm{e}}\xspace}
\newcommand{\freqc}{\ensuremath{f_\mathrm{c}}\xspace}
\newcommand{\epsilonzero}{\ensuremath{\epsilon_\mathrm{0}}\xspace}
\newcommand{\fgamma}{\ensuremath{f_\mathrm{\gamma}}\xspace}
\newcommand{\keV}{\kilo\electronvolt}
\newcommand{\eV}{\electronvolt}
\newcommand{\fW}{\femto\watt}

\begin{document}

\title{Single electron detection and spectroscopy via relativistic cyclotron radiation}

\newcommand{\uw}{Center for Experimental Nuclear Physics and Astrophysics, 
and Department of Physics, University of Washington, Seattle, WA, USA 98195}
\newcommand{\pnnl}{Pacific Northwest National Laboratory, Richland, WA, USA 99352}
\newcommand{\mitt}{Laboratory for Nuclear Science, Massachusetts Institute of Technology, Cambridge, MA, USA 02139}
\newcommand{\ucsb}{Dept.~of Physics, University of California, Santa Barbara, CA, USA 93106}
\newcommand{\kit}{Institut f\"{u}r Kernphysik, Karlsruher Institut f\"{u}r Technologie, Karlsruhe, Germany 76021}
\newcommand{\nrao}{National Radio Astronomy Observatory, Charlottesville, VA, USA 22903}

\newcommand{\comment}[1]{}

\affiliation{\pnnl}
\affiliation{\nrao}
\affiliation{\ucsb}
\affiliation{\uw}
\affiliation{\mitt}
\affiliation{\kit}

\author{D.\,M.~Asner}\affiliation{\pnnl}
\author{R.\,F.~Bradley}\affiliation{\nrao}
\author{L.~de Viveiros}\affiliation{\ucsb}
\author{P.\,J.~Doe}\affiliation{\uw}
\author{J.\,L.~Fernandes}\affiliation{\pnnl}
\author{M.~Fertl}\affiliation{\uw}
\author{E.\,C.~Finn}\affiliation{\pnnl}
\author{J.\,A.~Formaggio}\affiliation{\mitt}
\author{D.~Furse}\affiliation{\mitt}
\author{A.\,M.~Jones}\affiliation{\pnnl}
\author{J.\,N.~Kofron}\affiliation{\uw}
\author{B.\,H.~LaRoque}\affiliation{\ucsb}
\author{M.~Leber}\affiliation{\ucsb}
\author{E.\,L.~McBride}\affiliation{\uw}
\author{M.\,L.~Miller}\affiliation{\uw}
\author{P.~Mohanmurthy}\affiliation{\mitt}
\author{B.~Monreal}\affiliation{\ucsb}
\author{N.\,S.~Oblath}\affiliation{\mitt}
\author{R.\,G.\,H.~Robertson}\affiliation{\uw}
\author{L.\,J~Rosenberg}\affiliation{\uw}
\author{G.~Rybka}\affiliation{\uw}
\author{D.~Rysewyk}\affiliation{\mitt}
\author{M.\,G.~Sternberg}\affiliation{\uw}
\author{J.\,R.~Tedeschi}\affiliation{\pnnl}
\author{T.~Th\"{u}mmler}\affiliation{\kit}
\author{B.\,A.~VanDevender}\affiliation{\pnnl}
\author{N.\,L.~Woods}\affiliation{\uw}

\collaboration{Project 8 Collaboration}
\noaffiliation

\date{\today,\,\currenttime}

\begin{abstract}
It has been understood since 1897 that accelerating charges must emit electromagnetic radiation.  Although first derived in 1904, cyclotron radiation from a single electron orbiting in a magnetic field has never been observed directly.  We demonstrate single-electron detection in a novel radio-frequency spectrometer.  The relativistic shift in the cyclotron frequency permits a precise electron energy measurement. Precise beta electron spectroscopy from gaseous radiation sources is a key technique in modern efforts to measure the neutrino mass via the tritium decay endpoint, and this work demonstrates a fundamentally new approach to precision beta spectroscopy for future neutrino mass experiments.
\end{abstract}
\maketitle

For over a century, nuclear decay electron spectroscopy has played a pivotal role in 
the understanding of nuclear physics. Early measurements of the continuous $\beta$\hyp{}decay spectrum~\cite{Chadwick1914} provided the first evidence of the existence of the weak force and the neutrino~\cite{fermi_tentativo_1933}, and immediately hinted that the neutrino mass is small.  Continuing this tradition, present efforts to directly measure the mass of the neutrino rely on precision spectroscopy of the $\beta$\hyp{}decay energy  spectrum of \isotope{H}{3}.   Because the value of the neutrino mass is an input to the Standard Model of particle physics as well as precision cosmology, a precision measurement of the neutrino mass would represent a significant advance in our description of nature.  

The sensitivity of \isotope{H}{3}-based neutrino mass measurements has been improving over the past 80 years as a result of increasingly powerful electron spectrometry techniques~\cite{PhysRev.92.1521,Bergkvist1972371,Lobashev1985305,Picard1992345}. The most sensitive experiments to date place a limit on the electron-flavor weighted neutrino mass $m_\beta~\le$ \SI{2.05}{\eV}/c$^2$~at 95\% C.L. \cite{Kraus:2004zw,PhysRevD.84.112003,Olive:2014}:

\begin{eqnarray}
m_\beta^2 &=& \sum_{i=1,2,3}\left|U_{ei}\right|^2m_{\nu i}^2,
\end{eqnarray}

\noindent where the $m_{\nu i}$ are neutrino eigenmasses and $U_{ei}$ represent the elements of the Maki-Nakagawa-Sakata-Pontecorvo mixing matrix \cite{Olive:2014}.  A future experiment, the Karlsruhe TRItium Neutrino experiment (KATRIN)~\cite{angrik_katrin_2005}, is expected to be sensitive down to $m_\beta \ge$  \SI{0.2}{\eV}/c$^2$.  The absolute lower bound, which can be derived from the currently known neutrino oscillation parameters~\cite{Olive:2014}, is $m_\beta \ge$  \SI{0.01}{\eV}/c$^2$.

Since 1897, it has been known that accelerating charges emit
electromagnetic radiation~\cite{bib:Larmor}. Cyclotron radiation, the particular form of
radiation emitted by an electron orbiting in a magnetic field, was
first derived in 1904~\cite{Heaviside1904}.  Single electrons undergoing cyclotron motion in Penning traps have been previously detected non-destructively via image currents~\cite{PhysRevLett.38.310}, and relativistically-shifted cyclotron energy levels have been successfully utilized in precision measurements of the magnetic moment of the electron~\cite{PhysRevLett.59.26, PhysRevLett.100.120801}.  Yet, cyclotron radiation from single electrons has not been observed directly.  

Consider the ideal case where an electron is created in the presence of a
uniform magnetic field $B$.  The subsequent orbits of the electron have a cyclotron frequency \fgamma that depends on the kinetic energy $K$ of the electron:

\begin{equation}
\label{eq:cyclotron}
\fgamma \equiv \frac{\freqc}{\gamma} =
	\frac{e B}{2 \pi \gamma\me},
\end{equation}

\noindent where $e \left(\me\right)$ is the electron charge (mass), $c$ is the 
speed of light in vacuum, and $\gamma=\left(1+K/\me c^2\right)$ is the Lorentz 
factor. The nonrelativistic frequency \freqc is \SI{2.799249110(6)E10}{\hertz} 
at \SI{1}{\tesla}~\cite{RevModPhys.84.1527}. The orbiting electron emits coherent 
electromagnetic radiation with a power spectrum that is strongly peaked at 
\fgamma. Due to the $K$ dependence of \fgamma, a frequency measurement of this radiation is related to the energy of the electron, and thus provides a new form of non\hyp{}destructive spectroscopy.

A frequency-based technique has, in principle, the capability of overcoming many of the limitations imposed by traditional spectroscopic techniques used in direct neutrino mass experiments using tritium.  The most sensitive methods in use today suffer from the need to extract the $\beta-$decay electron for measurement, imposing a practical limitation on the size and density of the tritium source used.  Because the gas is transparent to cyclotron radiation, this limitation does not apply to the cyclotron radiation detection technique.  An additional advantage over traditional techniques is provided by simultaneous sensitivity to an entire energy region of interest with event-by-event energy reconstruction, rather than a stepped integrating method. Furthermore, the measurement reconstructs the electron energy spectrum with well established techniques for measuring frequencies and magnetic fields. Here we demonstrate a technique for electron energy spectroscopy that directly measures the cyclotron radiation from single electrons. This technique, hereafter referred to as Cyclotron Radiation Emission Spectroscopy (CRES), could allow a future generation of experiments access to neutrino masses below the sensitivity floor of current experiments~\cite{Monreal:2009za}.

%
%

In free space, the total radiated power $P$ is given by the Larmor formula~\cite{Jackson1998}:

\begin{equation}
\label{eq:power}
P\left(\gamma,\theta\right)=\frac{1}{4\pi\epsilon_\mathrm{0}} 
\frac{2}{3}\frac{e^4}{m_\mathrm{e}^2 c} B^2 \left(\gamma^2\hyp{}1\right)\sin^2\theta,
\end{equation}

\noindent where \epsilonzero is the permittivity of free space and $\theta$ 
is the pitch angle of the electron, defined as the angle between the momentum
 vector of the electron and the direction of the magnetic field. For an electron
 with an energy near the \SI{18.6}{\keV} endpoint of \isotope{H}{3}, 
 approximately \SI{1.2}{\fW} is radiated in a \SI{1}{\tesla} magnetic field at a pitch angle of $\SI{90}{\degree}$.
 
The Project 8 collaboration has constructed an experiment designed to detect the cyclotron radiation from single electrons.  At the heart of the experiment is a small volume hereafter referred to as the ``cell'', in which a gaseous radioactive isotope is present at low pressure. In a uniform magnetic field, electrons from decays inside the cell emit cyclotron radiation.  The cell consists of a section of rectangular waveguide sized to capture and transmit the microwave radiation to the input of a low-noise radio-frequency receiver and digitizer.

The radioactive isotope \isotope{Kr}{83{\rm m}} is a gamma-emitting
isomer of \isotope{Kr}{83} with a half-life of \SI{1.8}{\hour}, in which internal conversion produces
mono\hyp{}energetic electron lines with kinetic
energies of \SIlist{17830.0(5);  30227(1); 30424(1);  30477(1);  31942(1)}{\eV}, with line widths less than 3 eV~\cite{Picard1992zz}. The short-lived \isotope{Kr}{83{\rm m}} is supplied at a steady rate by decays of a \SI{74}{\mega\becquerel} source of the parent \isotope{Rb}{83} with a half-life of \SI{86}{\day}, adsorbed onto zeolite beads~\cite{venos2005}.  The krypton diffuses freely from the zeolite and uniformly fills the
experimental system, including the cell, while non\hyp{}evaporable getter pumps reduce
the total pressure of non-noble gases to $<$~\SI{10}{\micro\pascal}.   The \isotope{Kr}{83{\rm m}} 
concentration and flow are monitored by means of a silicon detector
that is exposed to the gas system but is outside the cell and magnet.

\begin{figure*}
\begin{center}
\includegraphics[width=0.80\textwidth]{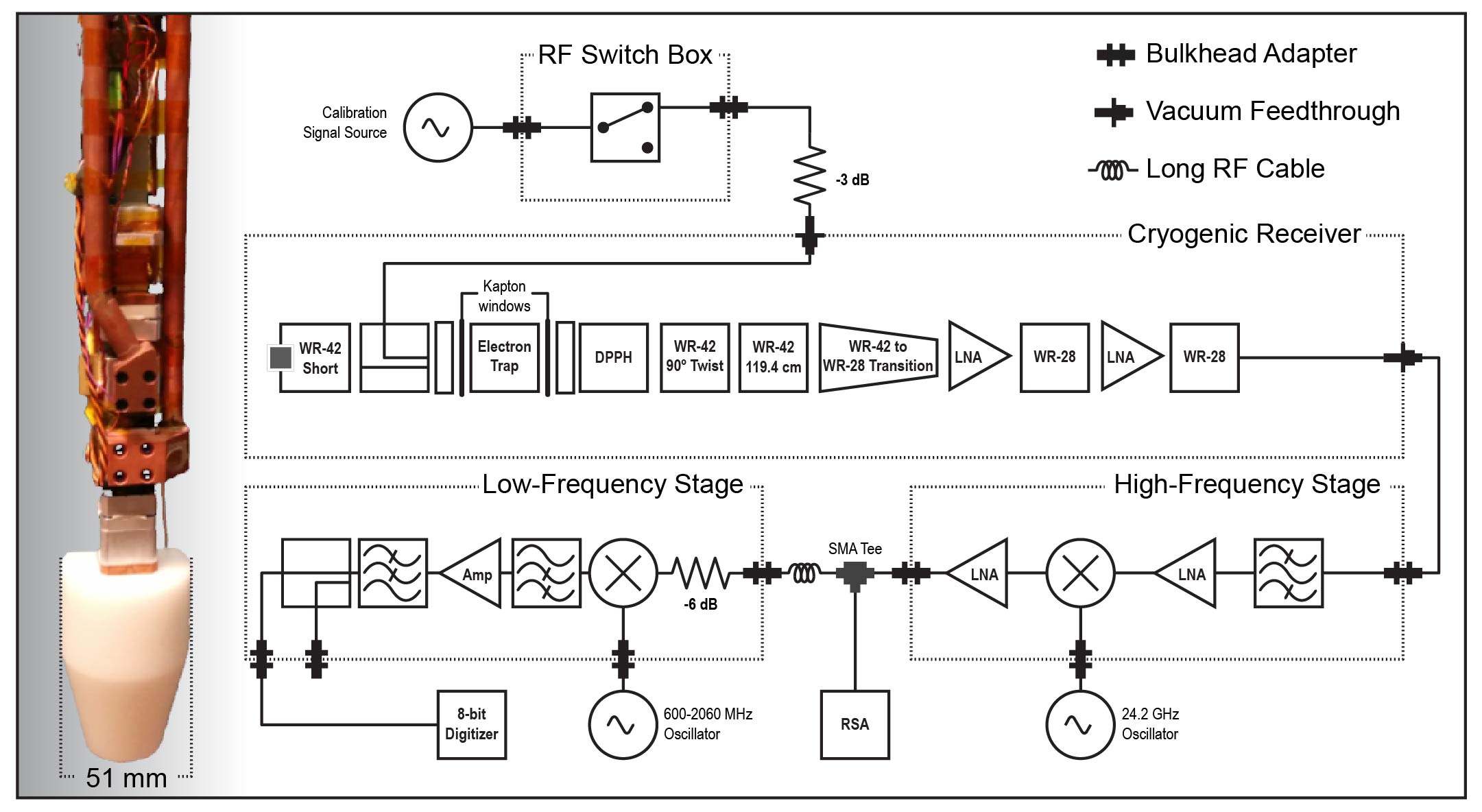} \\
\caption{A picture of the waveguide insert (left), where both the gas lines and 
trapping cell can be seen, and a corresponding schematic (right) of the 
receiver chain, consisting of cascaded cryogenic amplifiers, 
a high frequency stage and a low frequency stage. 
Calibrated radio frequency  signals can be injected into the cryogenic 
receiver and through the WR42 waveguide section. The high frequency band of 
interest, with a width of \SI{2}{\GHz} centered at \SI{26}{\GHz}, is then 
mixed down to be centered at \SI{1.8}{\GHz} with the same \SI{2}{\GHz} 
bandwidth.  A second amplification and mixing stage is used to further amplify the signal and shift the 
center frequency.  Data can be recorded either using an 8\hyp{}bit
digitizer or through a real\hyp{}time spectrum analyzer (RSA).}
\label{fig:receiver}
\end{center}
\end{figure*}

A magnetic field of approximately \SI{1}{\tesla} is provided by a \SI{52}{\milli\meter}\hyp{}diameter warm\hyp{}bore
superconducting solenoid magnet.  To allow sufficient time for detection and precise measurement of the emission  frequency, a weak magnetic trap is introduced at the midpoint of the cell to confine a small fraction of the produced electrons.  A  copper coil provides a near\hyp{}harmonic magnetic field perturbation with a gradient of up to \SI{100}{\tesla\per\meter\squared} along the magnetic field axis and a maximum depth of \SI{-8.2}{\milli\tesla} at the trap's center for an applied current of \SI{2}{\A}.   With the strongest trap settings, an electron that is emitted with a pitch angle $\gtrsim \SI{85}{\degree}$  will be confined until  collisions with the residual gas scatter the electron to a lower pitch angle.

The field strength inside the magnet bore is calibrated using both nuclear magnetic resonance (NMR) and electron spin resonance (ESR).  A full NMR map was used to assess the field homogeneity of the background field near the center of the waveguide cell.   ESR scans along the axis can be made with the waveguide in place by  observing absorption of microwaves by a sample of 2,2\hyp{}diphenyl\hyp{}1\hyp{}picrylhydrazyl (DPPH) inside a sealed Pyrex ampoule resting on the upper window of the cell.   The field strength at the center of the cell (without the additional trapping coil field) is measured to be \SI{0.9467(1)}{\tesla}.  Due to a slow drift in magnetic field and the effect of the trapping coil, the NMR and ESR measurements of the magnetic field were used only to predict the frequency region of interest and the achievable energy resolution.

At this field strength, the fundamental cyclotron signals for the \SI{30.4}{\keV} and \SI{17.8}{\keV} electrons
are expected to lie in the microwave K band.  Thus, the cell is constructed from a standard WR42 rectangular waveguide section ($10.7 \times 4.3$ \si{\milli\meter\squared}) coaxial with the solenoidal field. The \isotope{Kr}{83{\rm m}} source gas is confined to the \SI{7.6}{\centi\meter}\hyp{}long cell with \SI{25}{\micro\meter}\hyp{}thick Kapton windows. 

The motion of the electron in the cell couples strongly to the fundamental TE$_{10}$ mode of the WR42 waveguide, and most of the radiated power is emitted into this mode.  The remainder of the power is coupled to higher order modes, which are non\hyp{}propagating at the  cyclotron frequency of the electron and are therefore unobservable to the receiver. The cell is coupled to the receiver by a waveguide about \SI{1}{\meter} in length. The first stage of the receiver consists of two cascaded 22\hyp{}\SI{40}{\GHz} low noise preamplifiers that establish a noise floor of \SI{20(5)E-22}{\watt\per\hertz} referred to the cell.  The gain of this amplifier cascade is \SI{54}{\dB}, making the noise contribution from the components following these amplifiers negligible.  The frequency band of interest (from \SI{25}{\GHz}  to \SI{27}{GHz}) is mixed down with a local \SI{24.2}{\GHz} oscillator to a center frequency of \SI{1.8}{\GHz}.  A second mixer with a variable local oscillator frequency combines with a low\hyp{}pass filter to select a frequency subband of \SI{125}{\MHz} for narrowband signal analysis.  Signals are digitized at 250 mega-samples per second with a free\hyp{}running 8\hyp{}bit digitizer and recorded to disk.  A schematic of the receiver is shown in Figure~\ref{fig:receiver}.

The preamplifier performance is strongly dependent on physical temperature, 
which is reduced to \SI{50}{\kelvin} by a Gifford\hyp{}McMahon cryocooler.  At that  physical 
temperature, an equivalent noise temperature for the system may be derived by comparison with 
the available noise power from a matched resistor at temperature $T_e$,
which is $k_BT_e\,\si{\watt\per\hertz}$.  Converting the amplifier noise floor to an 
available noise power, the noise 
temperature of the receiver is roughly \SI{145}{\kelvin}.  The expected
signal\hyp{}to\hyp{}noise ratio (SNR) may be expressed as the ratio $P/k_B T_e \Delta f_\gamma$, where $\Delta f_\gamma$ is the bandwidth. For an \SI{18}{\keV} electron, the available signal\hyp{}to\hyp{}noise ratio is \SI{12}{\dB} for a receiver detection bandwidth of \SI{30}{\kilo\hertz}.

The presence of a magnetic trapping field shifts the cyclotron frequency. Approximating the trap as harmonic, the primary signal frequency is

\begin{equation}
\label{eq:receiver}
f(t) \simeq \frac{\freqc}{\gamma} 
\left(1 + \frac{\cos^2{\theta}}{2\sin^2{\theta}}\right)
\left(1 + \frac{P t}{\gamma m_e c^2}\right),
\end{equation}

\noindent where it can be seen that electrons with pitch angles deviating from $\theta = 90^{\circ}$ emit cyclotron radiation at a higher frequency because they explore higher magnetic\hyp{}field regions of the trap.  As the electron radiates energy, the frequency will increase.   Additional frequency structure is expected
due to the axial and magnetron motion of the electron in the trap,
but spectral lines due to this structure are relatively weak and have
not yet been observed.

Equation~\ref{eq:receiver} outlines the unique characteristics of a signal from a trapped
electron. One expects to find a nearly\hyp{}monochromatic RF 
signal at a frequency that slowly increases as the electron loses energy at a rate 
given by the Larmor formula.  Electrons that scatter off the gas at small angles may remain confined and continue to radiate power at a frequency determined by the new pitch angle and the energy lost in the collision. 

\begin{figure*}%
\centering
\includegraphics[width=0.90\textwidth]{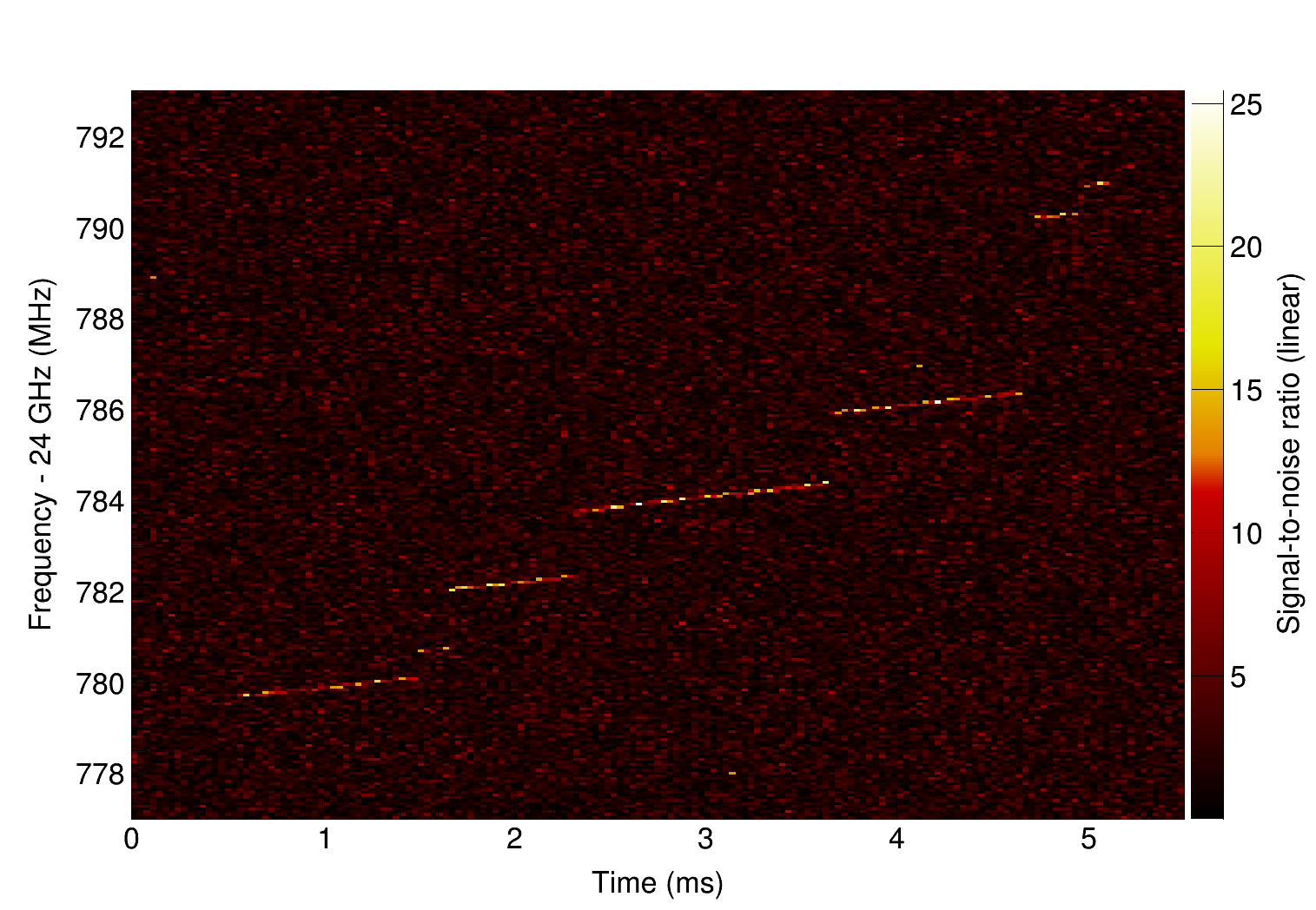}
\caption{A typical signal from 
the decay of \isotope{Kr}{83{\rm m}} characterized by an abrupt 
onset of narrowband power over the thermal noise of the system.  The measured frequency reflects the 
kinetic energy of the electron, in this case $30~\si{\keV}$. The frequency increases slowly 
as the electron loses energy by emission of cyclotron radiation, ending in the first of six or possibly seven visible frequency jumps before the electron is ejected from the trap.  The frequency-time window shown represents only a portion of an extended event lasting more than \SI{15}{\milli\second}. The sudden jumps result from the energy loss and pitch-angle changes caused by collisions with the residual gas, predominantly hydrogen.  The most probable size of the energy jump, as determined from many events, is \SI{14}{\eV}.}%
\label{fig:1stevent}%
\end{figure*}

A Short Time Fourier Transform, with a window size of 8192 samples, is performed on the digitized 
time series data. The resultant spectrogram has pixels with dimensions of \SI{30.52}{\kilo\hertz} by \SI{32.8}{\micro\second}.  Bins in the spectrogram that exceed the noise floor by \SI{8.12}{\dB} are then extracted. The resulting reduced set of 2D data is examined for structures which appear linear in the time\hyp{}frequency plane.  Linear segments which are so discovered are then grouped in time into fully reconstructed electron events. With the expected SNR, random alignments of power fluctuations should not occur.

The events recorded have precisely the 
characteristics outlined above.  Figure~\ref{fig:1stevent} shows the signal from a \SI{30}{\keV} 
electron observed during the first few milliseconds of data collection.  The features expected for electron
cyclotron emission are clearly evident, including (a) the abrupt onset of
narrowband RF power above the surrounding background, (b) a quasi\hyp{}linear
increase in frequency over time as the particle loses energy via cyclotron
emission and (c) sudden shifts in frequency due to gas collisions in which the electron remains magnetically confined. 

The power spectrum shown in Figure~\ref{fig:1stevent} is normalized to peak signal power. The Larmor prediction for free-space radiative loss at a 90-degree pitch angle is \SI{1.74}{\fW}.  The rate of change of frequency is a direct measure of the power radiated by the electron, and is expected to differ from the free-space prediction because the emitted power must couple into modes of the waveguide. For the electron shown, the rate of change in frequency of the longest duration track is measured to be \SI{1.61(4)}{\fW}.  The received power, \SI{0.66(16)}{\fW}, is \SI{3.9(10)}{\dB} below the radiated power owing to radiation into harmonics and axial sidebands and coupling to non-propagating waveguide modes.

\begin{figure}
\includegraphics[width=0.5\textwidth]{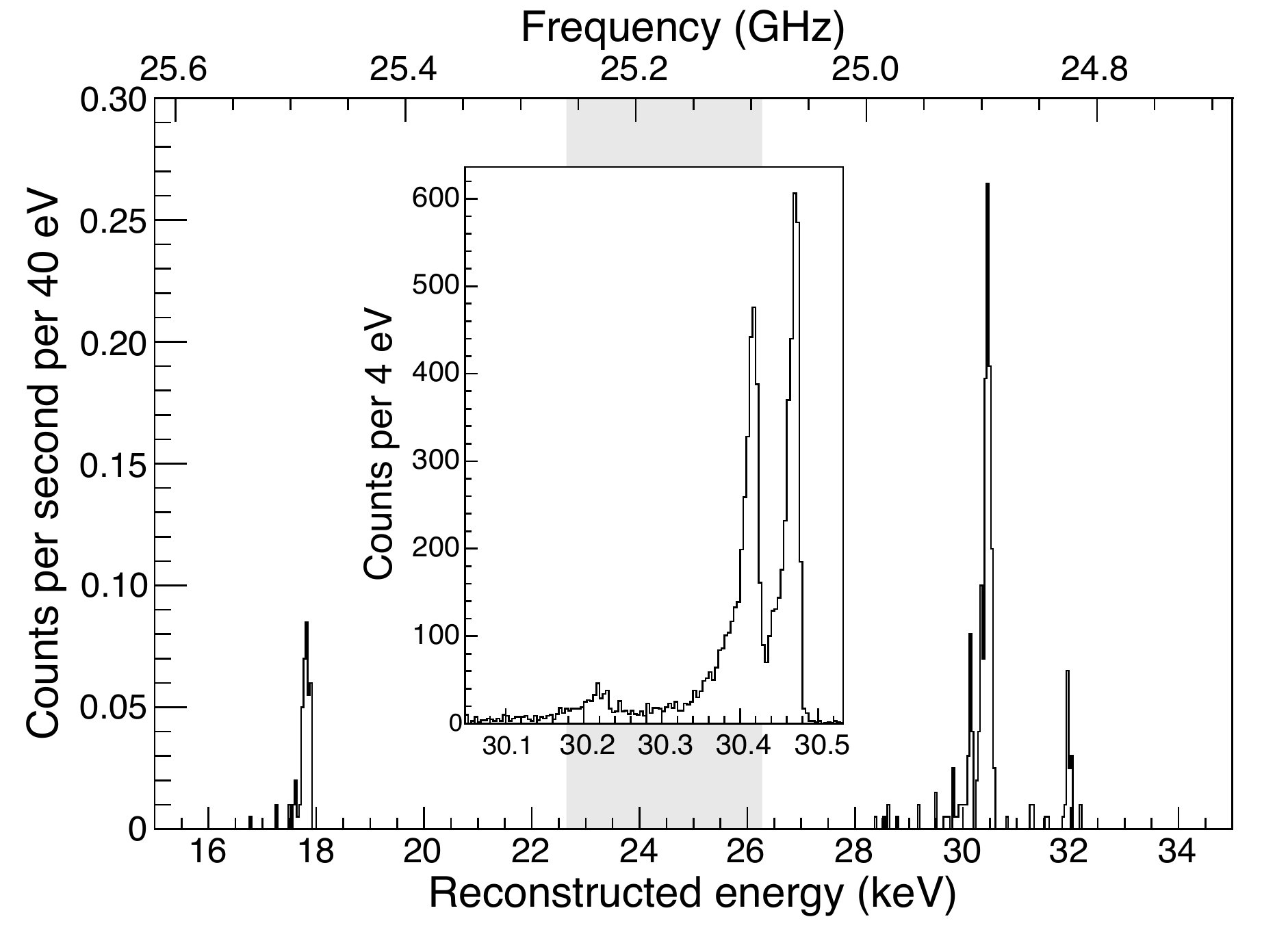}
\caption{The kinetic energy distribution of conversion electrons from \isotope{Kr}{83{\rm m}} as 
determined by CRES for a trapping current of \SI{800}{\mA}. The spectrum shows the \SI{17}{\keV}, \SI{32}{\keV} and \SI{30}{\keV}\hyp{}complex conversion electron lines. The shaded region indicates the bandwidth where no data were collected.  
Inset: With the trap current reduced to \SI{400}{\mA}, the feature at \SI{30.4}{\keV} is resolved into 3 lines.  Also visible as low-energy shoulders on these lines are shakeup satellites.}
\label{fig:Krpeaks}
\end{figure}

A clear excess of candidate events over background can be seen at \SI{17.8}{\keV}, \SI{30}{\keV} and \SI{32}{\keV} (see Figure~\ref{fig:Krpeaks}).  As a check for backgrounds, data collected with the trap de-energized were also analyzed. No events were identified under those conditions. The ratio of the  \SI{30}{\keV} to \SI{17.8}{\keV} peak frequencies, which is independent of the absolute magnetic field, is measured to be \num{1.023870(60)}, in very good agreement with the expected weighted average peak ratio of \num{1.023875(2)}.  The uncertainty in the magnetic field from the DPPH measurement corresponds to an absolute energy uncertainty of \SI{60}{eV}. Alternatively, one can use the \SI{17.8}{keV} electron emission line to calibrate the mean field probed by the trapped electrons, as has been done for Figure~\ref{fig:Krpeaks}. This calibration method yields a mean magnetic field of \SI{0.9421(3)}{\tesla}, with a relative uncertainty in the energy of about \SI{30}{eV} and is currently limited by statistics as well as prior knowledge of the spectral line shape. A fit to the frequency distribution using a skewed gaussian line shape~\cite{L'Hoir1984336} yields a full width at half maximum (FWHM) of approximately \SI{130}{eV} and \SI{140}{eV} for the \SI{17}{\keV} and \SI{30}{\keV} emission lines, respectively.  

Improved energy resolution is expected as the trapping field decreases because pitch-angle spread is reduced (Eq.~\ref{eq:receiver}).  To demonstrate this point, data have been collected with a trapping current of \SI{400}{\mA} (\SI{-1.6}{\milli\tesla} trapping field) using a real-time spectrum analyzer configured to trigger if the power in a \SI{10}{\kHz} bin exceeded a threshold \SI{2.5}{\dB} above a frequency-dependent noise floor.  Each trigger resulted in a 5 ms-long time series of phase-quadrature samples of a \SI{40}{\MHz}-bandwidth, including \SI{1}{\ms} prior to the trigger. The center of the bandwidth was alternately chosen such that either the \SI{17}{\keV} or \SI{30.4}{\keV} krypton emissions, and their associated excitations to higher energy bound states (shake-up) or to the continuum (shake-off), would be included~\cite{PhysRevLett.67.2291}.  The results, shown in Figure~\ref{fig:Krpeaks}(inset), illustrate the improvement in resolution, with the \SI{30.4}{\keV} doublet clearly resolved.  The resulting FWHM of the \SI{30.4}{\keV} lines is \SI{15}{\eV}, representing an order-of-magnitude improvement as compared to the \SI{800}{\mA} data.

The fundamental energy resolution achievable with this technique depends on two factors: uncertainty in measuring the emission frequency, and uncertainty in the time at which the emission begins.  Because of energy loss to radiation, the frequency is not constant, but increases quasi\hyp{}linearly with time.  The precision in the time of onset has a fundamental quantum limit and a practical limit from thermal noise. The inherent resolution will be further broadened according to the sampled field inhomogeneity, which dominates the energy resolution in our harmonic trap.

In summary, cyclotron radiation emission from single, mildly relativistic electrons has been observed experimentally.  The observation renders frequency-based measurements of electron kinetic energy, with the advantages of precision and independence from nuclear and atomic standards, a practical approach. An important and promising application for CRES is the measurement of the mass of the neutrino.

\section*{Acknowledgments}  The Project 8 collaboration acknowledges
financial support received from the University of Washington Royalty Research
Foundation, the Massachusetts Institute of Technology Wade Fellowship, the U.S.
 Department of Energy Office of Science, Office of Nuclear Physics to 
 the University of Washington under Award Number DE-FG02\hyp{}97ER41020, to the University of California, Santa Barbara under Award No. DE-SC0004036, and to the Massachusetts Institute of Technology under Award Number DE\hyp{}SC0011091, the National Science Foundation under Award Number 1205100, and the Laboratory Directed Research and Development Program at Pacific Northwest National Laboratory, a multiprogram national laboratory operated by Battelle for the U.S. Department of Energy under Contract DE-AC05\hyp{}76RL01830.  The Project 8 collaboration also wishes to thank: Dr. Otokar Dragoun and Dr.
Drahoslav Venos for providing us with the zeolite for our source;  Dr. Stefan Stoll for advice on ESR sources; and Dr. Jonathan Weintroub and Dr. Shep Doeleman for help with signal digitization development.   A portion of the research was performed using PNNL Institutional Computing at Pacific Northwest National Laboratory. The isotope(s) used in this research were supplied by the United States Department of Energy Office of Science by the Isotope Program in the Office of Nuclear Physics. \\

\bibliographystyle{apsrev4-1}
\bibliography{CRES_PRL_2014_revised}

\end{document}